# Multiband strain balanced superlattice material system for third generation infrared detectors

Arash Dehzangi, Member IEEE

Department of Electrical and Computer Engineering, Robert R. McCormick School of Engineering and Applied Science, Northwestern University, Evanston, IL, USA

*Abstract*— Recent increasing interest in strain balanced Type-II superlattices material causing close attention from industry. Tremendous investment was drawn toward establishing strain balanced superlattice (SLS) as new alternative for infrared photodetectors across the broad range of infrared spectrum. In recent years, SLS material system has shown capability in particular specifications to compete with mature and standard mercury cadmium telluride for mid-wavelength and long-wavelength infrared detection. It has been great interest in SLS material system for applications aligned with the standard of third generation of infrared detectors. In that level, photodetectors with multi-color detection capabilities based on SLS material system are highly desired. In this presentation, some recent progress in three color infrared photodetectors based on SLS material system will be presented.

## I. Introduction

Detection from multiple infrared bands to enhance the target discrimination and identification, is one of the requirements for third generation of infrared photodetectors. Multicolor photodetectors can offer both passive imaging (e.g. mid-wavelength and long-wavelength infrared, MWIR/LWIR) and active imaging short-wavelength infrared (SWIR).[1-4]

Multispectral detection is highly desired and became popular, where several two-color infrared focal plane array (FPA) imaging systems have been reported. Recently, photodetectors with three infrared bands have also been suggested.[5, 6] The goal is to incorporate multicolor functionality into on single detector, rather than compounding multiple single band detectors together.[7]

The first challenge is to find appropriate and relevant material for the task of multispectral detection with enough broad range of detection. Mercury cadmium telluride (MCT) and quantum well infrared photodetectors (QWIPs) are present available technology for two-color infrared detection. [8, 9] Given to the poor performance of QWIPs and low yield of MCT FPAs [7], III-V based Type-II superlattice material system with multispectral detection can be a viable choice. InAs/GaSb/AlSb SLS material with band gap engineering flexibility, high quality material with low defectivity and high uniformity appeared to be a promising material system for optoelectronic and photonic applications [10-13].

Another challenge for multi-color infrared detection is to provide low operating voltage and small gain-voltage slope for the infrared FPAs compatible to read out integrated circuits (ROICs), which in turn need to be compatible for specifications needed for multiband detectors.[14]. To meet the abovementioned requirements, low bias dependency is important for the multiband FPAs, meaning the bias at which the FPA reaches to its best optical functionality should be small enough to be compatible to ROIC's FET input capacitance and ROIC limitations. Low bias can support reasonable level of dark current to provide a high signal to noise ratio. [15].

Recently, new interest was given to triple-band photodetectors with three spectrally selective bands in the SWIR (2–3 µm), MWIR (3-5 µm) and LWIR (8–12 µm) regions. In this work some of recent progress in SLS triple-color photodetectors will be reviewed.[16], with the glance of future improvement in IR multiband detection and IR imaging.

## II. Multi-color Type-II superlattice infrared detectors based on InAs/GaSb

### A. Two color SLS infrared photodetectors

Two-color SLS infrared photodetectors are mostly based on InAs/GaSb or InAs/InAs$_{1-x}$Sb$_x$ material, grown by molecular beam epitaxy (MBE) on lattice match GASb substrate. In recent progress toward multiband infrared detection and imaging for SLS material, different novel device design and architecture were conveyed e.g., M-structure barriers [17]. High performance bias selectable two color FPA sensors SWIR-MWIR [18] and MWIR-LWIR [19], FPA sensors have been successfully demonstrated. It has been proven that devices with back-to-back p-i-n-n-i-p photodiodes came up with promising result, in which the signal in each channel in two-color has been collected independently by switching the polarity of voltage. (i.e., bias selectable [20]. This structure is simple and can operate with standard two contacts and using InAs/GaSb/AlSb heterostructure SLS for the absorption region and barriers allow optimization of both optical and electrical performance in each particular channels. In fact, realizing this dual-band back-to-back photodiode was the major step toward achieving the triple band SWIR-MWIR-LWIR SLS devices.

## B. Three color SLS infrared photodetectors

Tree-color infrared photodetector based on SLS material with two and three terminals for optical and electrical signal collection was recently reported. [5, 16] Technically three terminal stack imposes drastic fabrication difficulties for processing and losing valuable fill factor due to extra etching which in turn will reduce the quantum efficiency (QE). Therefore, two-terminal stack seems to be more reasonable fabrication approach, specifically when it comes to FPA fabrication and adjustability to available ROICs in market.

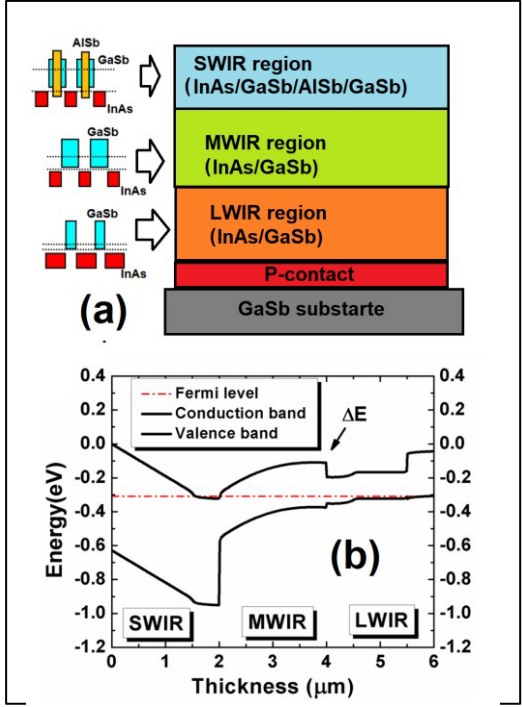

Fig 1. a) Schematic diagram of a three-color photodetedtor with SLS structure. The colored rectangles shows band gap of component materials. Dotted lines showing the actual band gap of the specifi superlatice dsign (b) Schematic of the band diagram of the device where the red dotted line representing the Fermi level,

SLS InAs/GaSb/AlSb based three-bands photodetector using two terminal stack and implementing M-structure has been reported. The structure, band diagram and SLS composition for each channel are shown in Fig 1. Each channel absorbs the radiation according to particular cut-off of each channel. SWIR absorption with the widest bandgap is located at the top of MWIR and the LWIR channel is located at the bottom of the structure. The design is bias selectable with two contacts, meaning that the collection of signals for each channel is managed by the polarity of the bias. The schematic band alignment in Fig 1b, large band off set between SWIR/MWIR (coming from M-structure design) acting as a barrier against the transport of the majority carrier holes, where the electrons as minority carriers can move with no impediment. This will suppress the cross talk between the two channels. On the other hand, the conduction band offset ($\Delta E$) between the LWIR/MWIR area can act as a barrier for electrons and control the voltage at which each channel will become functional (turn on voltage). The material was grown on n-type Tellurium (Te)-doped (001) GaSb substrate using a molecular beam epitaxy (MBE) equipped with group III SUMO® cells and group V valved crackers. The operation principle of the device, growth and structure details were explained in previous works [19]. By applying reversed bias on SWIR channel and only SWIR signal will be collected. Applying positive bias will activate MWIR and LWIR layers are in forward mode. Positive bias can activate the MWIR and LWIR channels and appropriate bias can extract desired signals from each channel.

For the design of SLS material and M structure as a barrier careful band structure engineering was performed using quantum mechanical modelling. For the band structure calculation and cut-off wavelength determination the empirical tight binding model (ETBM) was used [21, 22]. Using ETBM, the $\Delta E \sim 125$ meV was calculated. As an example, Fig 2 displays the calculated map for cut-off wavelength for InAs/GaSb/AlSb/GaSb. The calculation was performed based on thickness of AlSb and InAs, where the thickness named as monolayers (ML) for each component materials. Based on this map and according to acceptable range of lattice mismatch (the area between 3000 ppm and -3000 ppm based on MBE growth limitation shown in bold red contours), the desired SLS M structure and cut-off can be selected. The dashed curves show the equal contours of cut-off wavelength.

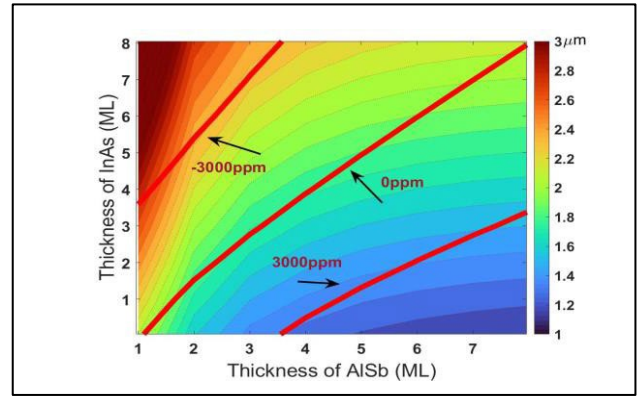

Fig 2. Calculated map of 50% cutoff wavelengths for the SLS InAs/GaSb/AlSb/GaSb basd on InAs and AlSb layer thickness with the unit of monolayer (ML).

It is worth nothing that careful band structure alignment and design is extremely crucial for the three-color device to make it operational based on polarity of the applied bias.

The optical performance and electrical perforamnc of the three-color photodetector device is shown in Fig 3. The optical characterization of the triple-band photodetector devices was performed in front-side illumination configuration with no applied anti-reflection (AR) coating. A Bruker IFS 66v/S Fourier transform infrared spectrometer (FTIR) was implemented to measure the spectral response. As it can be seen the optical respond is bias selectable and by changing the polarity and bias voltage different cut-off wavelengths are extracted. At 77 K (Fig 3a), the SWIR at -1.0 V bias showed 40% QE at ~1.5 µm, where MWIR and LWIR signal were extracted at +1.0 V and +4.5 V, with QE of 25% (at ~4.0 µm) and 19% (at ~1.5 µm), respectively. At 77 K, the 100% cut-off wavelength was 2.3 µm, 4.8 µm for and 8.6 µm for SWIR, MWIR and LWIR, respectively.

The QE of the device was calculated using a calibrated blackbody source at 1000 °C. There is an undesired spectral cross-talk effect occurs between the LWIR and MWIR channel. This is a limiting factor for the three-color detector especially at longer wavelength and must be controlled. It is an ongoing research on controlling the spectral cross-talk and different designs and architectures have been proposed to address the issue.

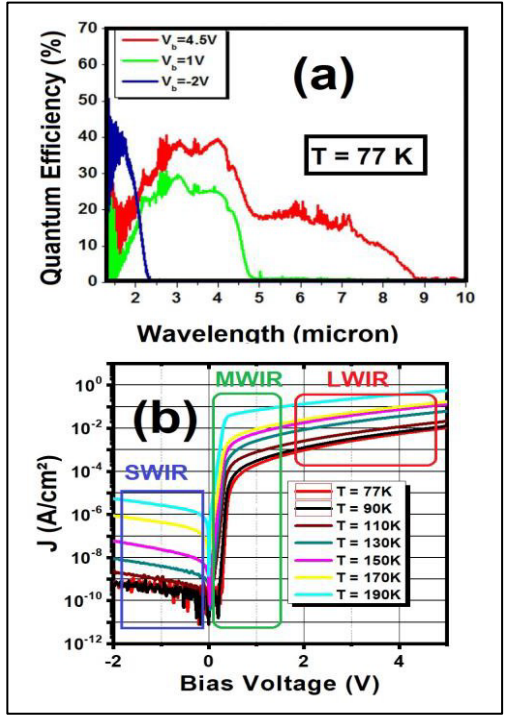

Fig 3. (a) QE spectrum of the SLS triple-color device at 77 K as function of applied bias. (b) Dark current of the device versus bias voltage at different temperature

To give a deeper view of the device performance the shot noise limited detectivity (D*) of the device was also calculated using following equation:

$$D^* = R_i \left[ 2qJ + \frac{4k_b T}{RA} \right]^{-1/2} \quad (1)$$

where T is operation temperature, $k_b$ is the Boltzmann constant, $R_i$ is the responsivity, J is the dark current density and $R_A$ is the differential resistance area product. Fig 4 shows the D* of the SLS infrared detector device at 77 K.

TABLE 1: The summary of D* for the triple-color SLS detector

|  | Temperature 77 K | | |
|---|---|---|---|
|  | *SWIR* | *MWIR* | *LWIR* |
| Bias Voltage | *-1.0 V* | *+1.0 V* | *+4.5 V* |
| D* (cm. $Hz^{1/2}$/W) | $3.0 \times 10^{13}$ | $1.0 \times 10^{11}$ | $2.0 \times 10^{10}$ |
| Wavelength (μm) | 1.7 | 4.0 | 7.2 |

The result for D* is summarized at Table 1 at different bias and relevant wavelength (λ) that the peak responsivity was calculated. The overall performance of the three-color SLS photodetector with two terminals in general is reasonable and better than reported three terminal devices. Optical signal is promising with distinctive three operational responses associated with three infrared bands. QEs for each channel is also reasonable, which can get better for LW/MW channels by increasing the thickness of the absorption regions and improving the etch profile and preserving the fill factor.

However, the device suffers from LW/MW cross-talk as explained earlier, but the important drawback is strong bias dependency and high bias voltage is needed to drive MWIR and especially the LWIR in operational regime. The 4.0 V value is not the bias voltage that can be accommodated by commercially available ROICs for the FPAs. Higher dark current is another concern which is not favorable for FPA applications. In this matter, to address the high bias dependency of the detector, a new approach was considered.

The structure was left untouched, and the focus was to improve the LW/MW operational regimes by incorporating non-uniform doping concentration across the MWIR active regions. The MWIR region was divided into two parts with different doping concentration ~ $10^{16}$ and p ~ $10^{15}$ cm$^{-3}$ with 1 μm thickness each, respectively. The schematic design of the doping level is illustrated in Fig 5 (more detail can be found in [23,24]). The total thickness of the device is still the same as previous design (6 μm).

The new doping level allowed to tune up the conduction band off-set and achieve better control on depletion region across the MWIR absorption region with lower bias dependency. In this nonuniform doping pattern, the lower doping level close to LW region and the higher doping is adjacent to SW region.

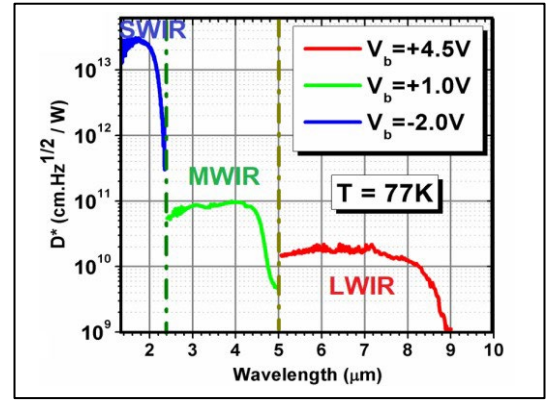

Fig 4. Detectivities of three-color SLS device at the wavelengths of interest at 77 K.

The goal is to modulate the conduction band off-set. The optical performance of the new design with nonuniform MWIR doping level supports the hypotheses and lowering the bias dependency of the three-colors color SLS photodetector.

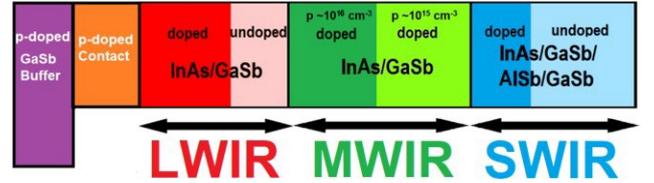

Fig 5. Schematic diagram of new structure of three-color SLS photodetector with nonuniform MWIR doping level

Fig 6a,b displays the result for optical performance of three-color SLS device with non-uniform doping level at 77K. It is clear that the bias dependency and turn-on voltage of the device has been improved across the MWIR and LWIR optical response. While the SWIR signal is kept unchanged for negative bias voltage, the MWIR response is now emerging in lower bias (0.5 V) and most importantly the LWIR signal saturation occurs at +1.5V.

The cut-off of the three regions is almost the same which implies a good control on growth and accurate band structure alignment. Saturated values for QE at peak responsivity are 37%, 25% (Fig 6a) and 20% (Fig 6b) for SWIR, MWIR and LWIR signal, respectively.

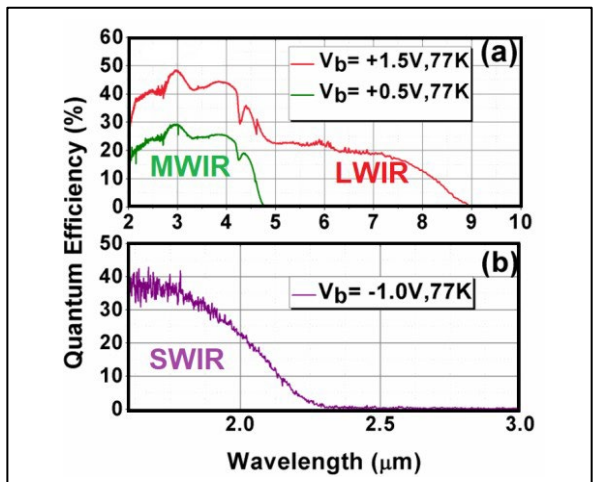

Fig 6 (a,b) QE spectra of three-color SLS device with non-uniform doping level at SWIR, MWIR and LWIR operational regime at different bias voltage (Vb = -1.0, 0.5 and +1.5 V)

## III. CONCLUSION

Recent advances in multi-band strain balanced Type-II superlattice photodetectors for third generation of infrared detection was reviewed. SLS is a viable alternative for future trend in infrared detection with low-dimensional quantum system. A triple-band SLS photodetector, based on InAs/GaSb/AlSb material for SWIR-MWIR-LWIR phot detection. The device designed using band structure engineering and quantum modeling and suggesting nonuniform doping level across the different active region to control high bias dependency. Latest result was reviewed and lower operating bias for the triple-band infrared photodetector was achieved.